\documentclass{aa}

\usepackage{graphicx}
\usepackage{txfonts}
\usepackage{color}
\usepackage{amsmath}
\usepackage{amssymb}
\usepackage{float}
\usepackage{url}
\usepackage{natbib}
\usepackage{ulem}
\usepackage{textcomp,gensymb}

\usepackage[utf8]{inputenc}

\usepackage{amsmath} % or simply amstext

\newcommand{\msun}{\mbox{$M_{\odot}$\,}}
\newcommand{\lsun}{\mbox{$L_{\odot}$\,}}
\newcommand{\kms}{\mbox{\,km\,${\mathrm{s}^{-1}}$\,}}

\newcommand{\mzams}{$M_{ZAMS}$}

\makeatother

\begin{document}

\title{Revisiting the progenitor of the low-luminosity type II-plateau supernova, SN~2008bk}
\author{D. O'Neill\inst{1}
   \and R. Kotak\inst{2}
   \and M. Fraser\inst{3}
   \and S. Mattila \inst{2}
   \and G. Pietrzy\'{n}ski\inst{4,5}
   \and J.~L. Prieto \inst{6,7}
   }
    \institute{Astrophysics Research Centre, School of Mathematics and Physics, Queen's University Belfast, Belfast, BT7 1NN, UK. \\ (\email{doneill955@qub.ac.uk} \label{inst1})
    \and Department of Physics and Astronomy, Vesilinnantie 5, University of Turku,  Turku FI-20014, Finland. \label{inst2}
    \and School of Physics, O'Brien Centre for Science North, University College Dublin, Belfield, Dublin 4, Ireland. \label{inst3}
    \and Nicolaus Copernicus Astronomical Centre, Warsaw, Poland. \label{inst4}
    \and Universidad de Concepci\'{o}n, Departamento de Astronom\'{i}a, Concepci\'{o}n, Chile.\label{inst5}
    \and N\'ucleo de Astronom\'ia de la Facultad de Ingenier\'ia y Ciencias, Universidad Diego Portales, Av. Ej\'ercito 441 Santiago, Chile. \label{inst6}
    \and Millennium Institute of Astrophysics, Santiago, Chile. \label{inst7}
    }
\abstract{

The availability of updated model atmospheres for red supergiants and improvements in single and binary stellar evolution models, as well as previously unpublished data prompted us to revisit the progenitor of low-luminosity type II-Plateau supernova, SN~2008bk. 
Using mid-IR data in combination with dust models, we find that high temperature ($4250-4500$\,K), high
extinction ($E(B-V)>0.7$) solutions are incompatible with the data.
We therefore favour a cool ($\sim3500-3700$\,K) progenitor with a luminosity of $\log(L/\lsun)\sim4.53$. Comparing with evolutionary tracks, we infer progenitor masses in the 8-10\,\msun range in agreement with some previous studies. This mass is consistent with the observed pattern of low-luminosity Type IIP SNe coming from the explosion of RSGs at the lower extremum for core-collapse.
We also present multi-epoch data of the progenitor, but do not find clear evidence of variability. 
}

  
\maketitle

\section{Introduction}

Although it is generally accepted that Type II Plateau supernovae (IIP SNe) arise from the core collapse of red supergiant (RSG) stars, inferring the mass range of progenitors that give rise to Type IIP SNe from observations is far more challenging.
The most direct way of detecting SN progenitors is by searching for these in
archival pre-explosion imaging. This necessarily involves an element of serendipity, and 
to date, only 13 progenitors of Type IIP/L SNe have been detected in this way; of these, 
only 6 have detections in more than two filters. 
Yet, type IIP SNe are the most frequently occurring core-collapse subtype \citep{LiRates}. 

Poor wavelength coverage, stemming from a lack of multi-band detections 
leads to a degeneracy between the inferred extinction and luminosity of the progenitor; these in turn result in large errors in the derived progenitor mass. At least two broadband colours sampling the optical to near-IR spectral energy distribution of the progenitor is desirable. 
Ideally, coverage of the progenitor's spectral energy distribution (SED) from the ultra-violet through to the infrared region would be available, but the bulk of archival data on hand is at optical wavelengths.

\begin{table*}[!t]
\renewcommand{\arraystretch}{1.1}
\begin{center}
\begin{tabular}{l c c c c c c l}
\hline\hline                 % inserts double horizontal lines
\tabularnewline[-0.25cm]
Distance & $E(B-V)_{\mathrm{tot}}$  & Temperature & Luminosity & Metallicity & Mass & Method & Reference \\
(Mpc)  & (mag) & (K) & $\log(L/L_{\odot}$) & $\log(Z/Z_{\odot}$) & \msun &  &  \tabularnewline% table heading 
\hline
3.91$\pm$0.41 & 0.32 & 3500$^{+150}_{-50}$ & 4.63$\pm$0.10 & -0.40 & 8.5$\pm$1.0 & SED+BC & \cite{2008bkVLT}\\
3.40$\pm$0.08 &  0.02 & 3600$\pm$50 & 4.57$\pm$0.06 & -0.22 & 8~--~8.5 & SED+BC & \cite{2008bkSVD}\\

3.8$^{+0.37}_{-0.33}$ & 0.67$\pm0.13$ & 4200$\pm200$* & 4.72$\pm0.14$ & -0.40 & 12$\pm2$ & SED+BC & \cite{Davies13}\\

3.5$\pm$0.10 & 0.77$^{+0.17}_{-0.21}$ & 4330$^{+330}_{-335}$ & 4.84$^{+0.1}_{-0.12}$ & -0.23 & 12.9$^{+1.6}_{-1.8}$$^\dagger$ & SED & \cite{2008bkMaund}\\
3.91$\pm$0.24 & -  & - & 4.53$\pm$0.07 & - & 8.3$\pm$0.6 & BC & \cite{IIpProDavies}\\
\tabularnewline[-0.25cm]
\hline
\tabularnewline[-0.25cm]
3.5$\pm$0.10 & 0.07$^{+0.09}_{-0.05}$ & 3500$\pm$160 & 4.54$^{+0.13}_{-0.09}$ & -0.25 & 8 (8-11)$^\ddag$ & SED & This study - \texttt{PHOENIX} \\ %4.57$^{+0.08}_{-0.05}$
3.5$\pm$0.10 & 0.88$\pm$0.09 & 4250--4500 & 4.88$^{+0.14}_{-0.06}$ & -0.25 & - & SED & This study - \texttt{MARCS} \\
3.5$\pm$0.10 & 0.02$^{+0.09}$ & 3500$\pm$150 & 4.51$^{+0.11}_{-0.05}$ & -0.25 & 8 (8-10)$^\ddag$ & SED & This study - \texttt{MARCS} \\

\hline\hline %inserts single line
\tabularnewline[-0.25cm]
\end{tabular}
\caption{Summary of previous estimates for the properties of the  progenitor star.  
*Models restricted to $T=4130\pm150$\,K.
$^\dagger$ Matched to the end of He burning phase of stellar evolution models.
$^\ddag$ Allowing for the full distance range (see also Fig. \ref{STARS}).}

\label{Mtable} 
\end{center}
\end{table*}

Early studies of directly detected progenitors found the zero-age main sequence mass (\mzams) range of stars producing type II SNe to be $8.5^{+1.0}_{-1.5}\lesssim\msun\lesssim16.5^{+1.5}_{-1.5}$ \citep[][and references therein]{RSGP}, which was markedly discrepant at the upper mass end with theoretical expectations \citep[e.g. $9\lesssim \msun \lesssim40$][]{Heger}. 
While the cause of this discrepancy has been discussed frequently in literature \citep[e.g.,][]{Smartt15,IIpProDavies,DaviesRSGP,KochanekRSGP}, 
the lower mass limit is also of interest from an evolutionary standpoint.
The (Fe) core-collapse of low mass RSGs is thought to give rise to low luminosity IIP SNe i.e., those with faint plateau magnitudes ($-13.5<M_V<-16$), low ejecta velocities \citep[$<$5000\kms][]{2005csPast}, and $^{56}$Ni masses as low as $\lesssim$0.01\msun \citep[e.g.,][]{97D}. 
However, similar properties may also arise due to the collapse of an O-Ne-Mg core in super-AGB stars with masses of $\sim9$\msun, giving rise to so-called electron-capture SNe \citep[e.g.,][]{ECSNPoel,Pumo09}.
Scenarios invoking higher mass RSGs have also been proposed for low luminosity IIP SNe. These generally require large amount of fallback to match the above properties \citep{Zamp03,Nomoto13}.
 Reliably distinguishing between these possibilities remains one of the challenges of SN research.

The low-luminosity and normal IIP SNe form a continuous distribution in brightness; 
here we simply adopt an upper brightness cut-off of $M_{\mathrm{visual}}\sim-16$.
Although more than a dozen low-luminosity IIP SNe have been classified as such, 
only 3 have  progenitor detections: SNe 2005cs, 2008bk, and 2018aoq
\citep{05csPro,Pro05csLi,2008bkVLT,18aoqPro}, respectively. 
SN~2005cs occurred in M51 (8.4\,Mpc); a progenitor was identified in archival {\it Hubble Space Telescope} (HST) images in only one filter (F814W). Combining this detection with upper-limits from  other filters, the progenitor mass was inferred to be 9$^{+3}_{-2}$\msun. 
SN~2018aoq occurred in NGC~4151 (18\,Mpc). With progenitor detections in both optical (F350LP, F555W, and F814W), and near-infrared (F160W) {\it HST} filters, it was estimated to have a mass of $\sim$10\msun \citep{18aoqPro}.

SN~2008bk merits special attention as it occurred in NGC~7793 at distance of only $\sim$3.5 Mpc \citep{2008bkDist,2008bkdist2}. A wealth of pre-explosion archival data ranging from
$3800-24000\,\AA$ were available, and the progenitor was clearly detected in multiple filters. A first analysis reported a progenitor mass of $\sim$8--9\msun \citep{2008bkVLT, 2008bkSVD}, which was later revised to $\sim$13\,\msun \citep{2008bkMaund}, suggesting that higher mass RSGs could give rise to low-luminosity IIP SNe. Here we revisit this issue, motivated in part by updated model grids, and in part by the existence of previously unpublished data of the progenitor.

\subsection{Previous studies of 2008bk}
\label{PrevStudies}

Analysis of the progenitor of SN~2008bk by \cite{2008bkVLT} and  \cite{2008bkSVD} showed that its SED could be fit by a \texttt{MARCS} \citep{MARCS}\footnote{Model Atmospheres in Radiative and Convective Scheme; https://marcs.astro.uu.se/index.php} model atmosphere with a temperature of $\sim$3500--3600\,K. Despite differences in the photometry and adopted extinction values  ($E(B-V)=0.32$ and $E(B-V)=0.02$, respectively), both studies infer a progenitor mass of $8.5\pm$1.0\,\msun\, (Table \ref{Mtable}).

Using late time imaging, \cite{2008bkDisap} found that the source previously identified as the progenitor had vanished, confirming its association with SN~2008bk. Using the photometry in \citet{2008bkVLT}, \citet{Davies13} showed that a higher temperature of T=4200\,K and extinction $E(B-V)=0.67$ also matched the progenitor SED.

 \citet{2008bkMaund} noted that previous attempts to characterise the progenitor, especially from ground-based imaging, were probably affected by the combination of a crowded field and patchy background. They therefore obtained late-time template images using identical instrument and filter combinations as the pre-explosion imaging, allowing them to remove contaminating flux. Based on Bayesian fitting of the template-subtracted progenitor photometry to \texttt{MARCS} models, they inferred a temperature of 4330\,K and $E(B-V)=0.77$. 
 This results in a significantly more luminous progenitor compared to previous studies, and a correspondingly more massive RSG, 12.9$^{+1.6}_{-1.8}$\,\msun at the end of its He burning phase. Interestingly, the extinction towards field stars in the vicinity of SN~2008bk was found to be only $E(B-V)=0.09$, leading \citet{2008bkPop} to conclude that their higher value must arise from CSM dust.

A recent study by \cite{IIpProDavies} favours a lower value of 8.3$\pm$0.6\msun on the grounds that RSGs evolve to later spectral types, and that this must be taken into consideration so that the appropriate bolometric correction can be applied. 
Radiative transfer and hydrodynamical modelling of the photospheric and nebular phases of SN~2008bk allow for a 9$-$12\msun progenitor \citep{2008bkLisakov,AndersLowMass,Martinez20}, thereby encompassing all previous progenitor mass estimates.

There are several estimates for the distance to 2008bk: a shorter 3.5\,Mpc estimate  derived from Cepheid variables and the tip of the red giant branch (TRGB) \citep[][respectively]{2008bkDist,2008bkDistTRGB}, and a longer 3.9\,Mpc estimate also via the TRGB \citep{2008bkDistTRGBold}. Following \cite{2008bkMaund}, we take the mean of the two more recent studies, that also favours the direct distance determination methods i.e., we adopt 3.5$\pm$0.1\,Mpc as the distance to the NGC~7793, unless stated otherwise (Table \ref{Mtable}). 

\section{Methods}

\subsection{Archival Data}

Motivated by the discrepancy between the inferred temperature and mass for the progenitor we 
report on a new analysis of the progenitor properties. For this, we used the late-time difference image photometry 
reported in \citet{2008bkMaund} as these measurements should suffer from minimal contamination in most filters. The pre-explosion images were taken across 3 epochs (2001, 2005, and 2007), while the templates are from 2011, using the same or similar instrumental set-up.

No templates free of SN emission were available for the $g'$ and $r'$ bands.
The difference between the pre-explosion and template-subtracted images is $<$0.2 mag for all filters except the $i'$ and $I$ bands. For the latter, the progenitor is fainter in the difference images by $\sim$0.4\,mag, suggesting field contamination.

We also present {\it Spitzer}/IRAC observations of the progenitor of SN~2008bk. Although there are 2 epochs of IRAC channels 1-4 pre-explosion imaging taken in 2004 by the SINGS programme (Table \ref{IRACtable}), they are separated only by a day. All post-explosion data is from the warm {\it Spitzer} mission. 
As noted previously, the SN occurred in a relatively crowded region, so template subtraction is a necessary step, especially given the IRAC pixel scale ($1\farcs2$/pix). 

We picked the 2018 Apr. 06 epoch as the reference. This choice is arbitrary, since the post-explosion data are taken between 2014--2019 i.e., long after the SN had faded. We checked that equivalent results were obtained by choosing a post-explosion epoch that was closest in orientation to the 2004 one. To carry out the subtraction, we used the post-basic calibrated data products. The image matching and subtraction was performed as implemented in the ISIS v2.2 image-subtraction package \citep{Alard98,Alard00}. 
The pre- and post-explosion images, and the difference images for both channels 
are shown in Figure \ref{ch1}. 

An aperture with a radius of 4 pixels was placed on the progenitor and its position was offset by 2-4 pixels in random directions around the progenitor position 9 times to ensure that the aperture position does not strongly dictate the results. 
We applied the appropriate aperture corrections to the average value resulting from the above procedure, and converted to the Vega magnitude system.

Although no templates are available for chs. 3 and 4, we can place limits on the progenitor brightness at these wavelengths.
We note that there is no significant flux at the progenitor location (Fig. \ref{ch34}). \citet{2008bkSVD} also noted the lack of 8.0\,$\mu$m (ch. 4) emission, and attributed this to low extinction. 
In order to estimate the depth of these images, we constructed approximate point spread functions using 
field stars in the image. The brightness of these artificial sources was adjusted until they were detected at the 3$\sigma$ level. 
The resulting photometry is given in Table \ref{IRACtable}.

\subsection{SED fitting}
In this study we use two different atmospheric model sets.
The MARCS model grid used here is based on spherically symmetric calculations for a 15\,\msun RSG star with surface gravities $-0.5\leq\log(g)\leq+1.0$, metallicities $-1.0\leq\log(Z/Z_\odot)\leq+0.5$, and a temperature range of 3300--4500\,K, with a step size of 100\,K for models below 4000\,K and 250\,K for those above this value.

The \texttt{PHOENIX} models \citep{Phoenix07} are also spherically symmetric and are part of the \texttt{STSYNPHOT} python package, which allows us to easily generate SEDs with a wide range of temperatures, metallicities and surface gravities. Here we adopt the same temperature range as for the \texttt{MARCS} models (3300--4500\,K), with a step size of 100\,K.

Many previous studies have pointed out that it is difficult to accurately capture the physical processes 
occurring in the outer layers of RSGs without computationally expensive detailed 3D hydrodynamical modelling \citep[e.g.,][]{Chiavassa09}. 
Nevertheless, the grids of 1D spherically symmetric models computed under the assumption of local
thermodynamic equilibrium have been shown to be in good general agreement with the optical-infrared SEDs \citep[e.g.,][]{Levesque05,Davies13}

   \begin{figure*}[!t]
   \begin{minipage}[l]{5in}
   \includegraphics[height=0.5\hsize]{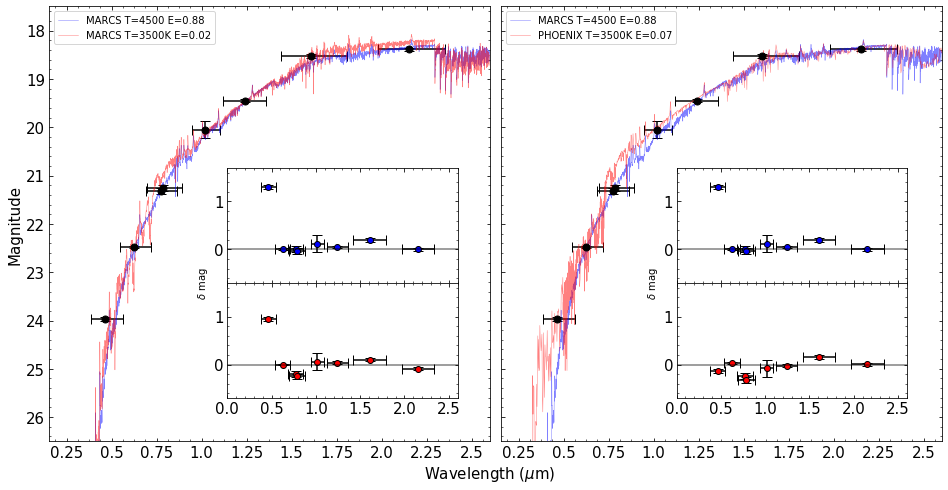}
   \end{minipage}
    \begin{minipage}[l]{2.2in}
      \caption{\textbf{Left:}Comparison between the two best fitting \texttt{MARCS} models. Shown in black points are the photometric points from the progenitor. The horizontal uncertainties simply denote the width of the photometric band. Inset are the residuals between the photometry and the SEDs for both fits. Note the large disagreement between the SEDs and the $g'$ band point at 0.46$\mu$m. \textbf{Right:} Comparison between the high temperature, high extinction \texttt{MARCS} model and the best fitting \texttt{PHOENIX} model. The \texttt{PHOENIX} model appears to fit the blue $g'$ band much better, however it overestimates the flux in the optical $>6500\AA$. The mid-IR points (Table \ref{IRACtable} are not shown here for clarity and ease of comparison with previous studies.}
         \label{MARCS}
         \end{minipage}
   \end{figure*}

Using a Monte-Carlo weighted fitting technique we attempted to match the model SEDs to the progenitor photometry. The initial fitting parameters were temperature, luminosity, line-of-sight extinction, for which we adopted $R_V=3.1$, and circumstellar medium (CSM) extinction, for which we use the equations in \citet{12awCSM}. We set a lower bound for the former, corresponding to the line-of-sight Milky Way contribution of $E(B-V)$=0.02 mag. \citep{SF2011}. 
We only considered models with a surface gravity value of $\log(g)$=0.0. All previous studies found that the metallicity of the progenitor environment was most likely sub-solar (see Table \ref{Mtable}), so we adopted models computed with a metallicity closest to previously reported values i.e., $\log(Z/Z_{\odot})=-0.25$.

The temperature, luminosity, and extinction parameters were sampled from a normal distribution with an initial central value; the initial temperature was set to 3900\,K, the initial brightness to $m_{r'}$=22\,mag, the line-of-sight extinction $E(B-V)$=0.3\,mag and the optical depth of the CSM  $\tau_V$=1, and allowed to vary between $0\leq\tau_v\leq20$.
We set the distributions to be wide enough to allow the entire range of parameters to be sampled. 
We then constructed SEDs from a randomly selected combination of temperature, luminosity, and extinction values, and computed the scatter between synthetic and observed magnitudes. If the scatter decreased compared to the previous best fitting iteration, the central values of the distribution were shifted to these, and the process repeated with the sampling width slowly narrowing if no improved fits are found. In effect, this results in the sampling distributions becoming narrower while moving through the parameter space. This process was carried out for 10000 iterations with the parameters producing the best fitting SED returned at the end of the process. The width of the sampling function at the end of this process is taken to be the resulting uncertainty provided that it is larger than the grid step size.

\section{Results \& Discussion}
\label{sec:models}

For the \texttt{MARCS} models, we found two combinations of parameters that matched the observed SED. The first resulted in a scatter $\delta_{mag}=0.14$~mag, and is a 3500$\pm$150\,K model with minimal extinction $E(B-V)\approx0.02$, similar to \cite{2008bkVLT} and \cite{2008bkSVD}. However, a similar match ($\delta_{mag}=0.15$~mag) is obtained with a 4500$\pm$250\,K model with a relatively high extinction of $E(B-V)=0.88$ that echoes the result reported by \cite{2008bkMaund}. The SEDs and residuals for both models are shown in the left-hand panel of Fig. \ref{MARCS}.\

While both SEDs match the majority of the optical and near infrared data points, there is a very large discrepancy between either model SED, and the $g'$ band magnitude ($\lambda_c=0.46\mu$m) resulting in a large ($\sim$1 \,mag) residual as shown in the inset panel of Figure \ref{MARCS}. As mentioned previously, the $g'$ band magnitude may well be contaminated by nearby unrelated sources. However, it is difficult to attribute such a large discrepancy solely to contamination because the $r'$ band point -- which also lacks a template -- is consistent with the model SED, implying that any contaminating source would emit predominantly at bluer wavelengths. 
We find no significant difference between SEDs with and without CSM extinction. 

\begin{figure}[!t]
   \centering
   \includegraphics[height=0.4\textwidth]{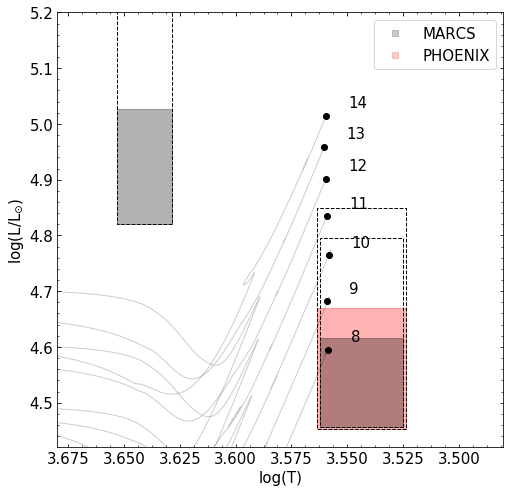}
      \caption{ Figure showing a comparison between stellar evolution tracks and the allowed regions
      as inferred from the observations. 
     Shown are the stellar evolution tracks of $8\le M_{ZAMS}/\msun \le14$ for sub-solar metallicity ($log(Z/Z_{\odot})=-0.4$).
      The results from this study are depicted by boxes that show the allowed ranges in temperature and luminosity for the progenitor given the uncertainties. The grey and red shaded regions show the luminosity ranges estimated using the SEDs of the MARCS and PHOENIX models, respectively, and 
      assuming the shorter (Cepheid and TRGB)
      distance of 3.5$\pm$0.1~Mpc for our two best model matches (\S \ref{sec:models}). 
      The dashed lines extend the boxes to show the effect of considering the full distance range ($3.32-4.32$\,Mpc).
      }
         \label{STARS}
   \end{figure}

We note however, that there is some degeneracy between the line-of-sight and CSM extinction in the sense that while the total visual extinction remained the same, the relative contribution of the components changed. We can infer the luminosity of the progenitor by correcting for extinction and integrating over the model SEDs. %in flux space. 
By comparing these results with STARS stellar evolution tracks  \citep{STARS2}, 
we can estimate the mass of such a star. For the hotter \texttt{MARCS} model, we find a luminosity of $\log(L/L_{\odot})=4.88^{+0.14}_{-0.06}$ but no clear match to any stellar evolution tracks. For the lower temperature one, we find a luminosity of $\log\left(L/L_{\odot}\right)=4.51^{+0.11}_{-0.05}$ that is consistent with the evolution of 8$-$9\msun stars. The region also includes the point at which the lowest mass (8\msun) model terminates \citep[see Table \ref{MARCS}][]{2008bkSVD}. Barring uncertainties in the models, we consider the effect of allowing a larger error in the distance estimate, (Fig. \ref{STARS}). 
For the \texttt{PHOENIX} models, we found the best match ($\delta_{mag}=0.06$~mag) with a $T=3500\pm160$\,K model and extinction $E(B-V)=0.07$, corresponding to $\log(L/L_{\odot})=4.54^{+0.13}_{-0.09}$, and matching the end point luminosity of an 8\msun track. 
We examined the impact of changing the metallicity to $\log(Z/Z_\odot)=-0.5$ and $0.0$ for both  \texttt{MARCS} and \texttt{PHOENIX} models, and found results that were consistent with the above.

We also compare our results to the BPASS binary stellar evolution models
\citep{BPASS2-1}. For the lower temperature MARCS and PHOENIX models, we find two groups
of binary models that match. The majority consist of an 8.5\msun star in a long period orbit ($\log(P_{\mathrm{days}})\gtrsim3$) that undergoes little or no interaction with its companion, essentially evolving as a single star.
The second grouping consists of mergers of lower mass stars with combined masses in the $6-9$\msun range. We were unable to find any matches with the higher temperature models. We can rule out the former given that no source was detected in deep late-time imaging \citep{2008bkDisap,2008bkMaund}.

As can be seen in the right-hand panel of Fig. \ref{MARCS}, the scatter of the \texttt{PHOENIX} model is smaller ($\delta_{mag} = 0.06$) compared to the \texttt{MARCS} models. In particular, the match to the $g'$ band point is significantly better, although there is a small excess between 0.75--1.5$\mu$m that is not apparent in the \texttt{MARCS} models.
Previous studies on RSGs have also noted the disagreement between \texttt{MARCS} and \texttt{PHOENIX} models at bluer wavelengths e.g.,
\citet{Plez} who noted a large ($<0.25$\,mag) difference for the $U-B$ colour for $T_{\mathrm{eff}}<4000$\,K. Given the complexity of these models, the source of this mis-match remains to be securely identified.

Many studies have noted the tension between the extinction inferred from
e.g. SED fitting versus that expected from studies of known RSGs \citep[e.g.,][]{12awCSM,RSGPdust}. Taken at face value, our findings support an 8-9\,\msun progenitor. 

However, detailed modelling of the SN allows a wider range of $\sim9-12$\,\msun, suggesting that it may only be possible to distinguish more finely between these alternatives following improvements in the input physics of stellar evolution codes.

The low extinction value for our cooler model appears to be at odds with the values typically inferred for RSGs \citep[e.g.,][]{Davies13}. One possibility is that the complexities of the dust properties are not wholly captured by the methods used above. Interestingly, most previous studies that report RSG progenitors of type IIP SNe, also report extinctions that are significantly lower compared to RSGs, but the extinctions values are often inferred from SN observations that obviously cannot constrain progenitor CSM dust properties, and indeed some or all of the pre-existing dust may be destroyed by the explosion.

We are thus faced with a choice of either accepting our cooler model, or making use of the 
common knowledge of the properties of RSGs, and insisting that the higher extinction solution is preferable. However, there is a further test that we can carry out: 
we can check whether the mid-IR photometry is consistent with the high extinction case, in the knowledge that RSGs are strong emitters at these wavelengths. In order to do so, we used the \texttt{DUSTY} code \citep{DustyManual} using the same parameters as in \citet{12awCSM} i.e., 50/50 silicate and graphite dust in a shell with radius R$_{in}/R_{out}$=2, and a dust temperature of $T_D$=1000\,K. We then applied this formulation of the CSM reddening to \texttt{MARCS} models with temperatures $3300-4500$\,K that had been previously reddened only with a line-of-sight component of $E(B-V)=0.05$. 

We then utilised a Markov Chain Monte Carlo fitting algorithm \citep[python package {\it emcee\/},][]{emcee} to determine the most likely temperature, CSM dust extinction and luminosity parameters based on the $0.6-8$\,$\mu$m SED. In keeping with \citet{2008bkMaund}, we did not include the $g$-band point in the fit (see also Fig. \ref{MARCS}). We find a solution with $T=3700^{+170}_{-110}$\,K, and $\tau_V=0.45^{+0.33}_{-0.27}$ that is a good match to the progenitor SED.
 
\begin{figure}[!t]
   \centering
   \includegraphics[width=0.4\textwidth]{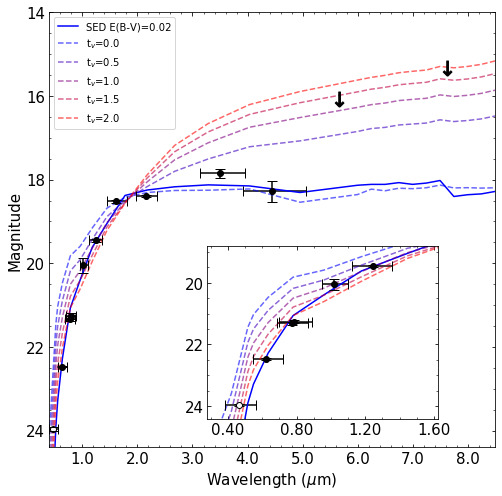}
      \caption{Output from the \texttt{DUSTY} models showing the hotter ($T=4500$\,K) \texttt{MARCS} model 
      with various amounts of CSM reddening (dashed lines) scaled to match the $H$ and $Ks$ bands. Also shown is the best fitting cooler ($T=3500$\,K, $E(B-V)$=0.02) model (solid line). Both have been re-binned to match the wavelength grid of the \texttt{DUSTY} models.
      The inset shows zoomed-in view of the optical region only; the open circle denotes the $g$-band point. It is immediately apparent that the mid-IR data is consistent only with low extinction. The low extinction, high temperature SED is too blue and does not match the optical data, while the high extinction, high temperature SED results in mid-IR emission that is too bright compared to the data. }
 
         \label{DUSTY}
   \end{figure}

A clear outcome is that when the dust emission is accounted for, the measured mid-IR magnitudes are significantly fainter than expected for a source with $\tau_v\sim2-3$. 
High temperature models with low CSM extinction are much too blue, while those with high extinction are much too bright in the mid-IR. These outcomes are depicted in Fig. \ref{DUSTY}. This exercise therefore favours a cool ($T=3500-3700$\,K), low extinction source. We checked that this outcome is robust against variations in dust composition and temperature.
  
\section{Progenitor variability}

Many previous studies have remarked upon an apparent temporal coincidence between eruptive mass loss and core-collapse, especially within the context of interacting (type IIn) SNe \citep[e.g.,][]{Smith14}. 
On the observational side, such a correlation is difficult to pin down given the inherent biases resulting from gaps in the data and depth of previous and current transient surveys. On the theoretical side, a number of processes have been identified that could give rise to outbursts prior to core-collapse \citep{Quataert12,SmithArnett14}, but it is not yet clear which of these processes, if any, occur in reality, and have observable consequences.
 
\citet{16fq} and \citet{Johnson18} discuss this specifically for the case of RSGs. Only 4 type IIP SN progenitors have multi-epoch observations: 
SNe~2013ej \citep[5 epochs over 5 years,][]{2013ejPro}, 2016cok \citep[15 epochs over 8 years,][]{16fq} and 2018aoq \citep[9 epochs over 1 year,][]{18aoqPro}, were observed in multiple optical {\it HST} filters with the progenitor of 2018aoq also having very limited {\it HST\/} IR data. The progenitor of 2017eaw has limited observations in the optical filters, but has an extensive mid-IR coverage \citep[34 epochs over 13 years,][]{2017eawPro}. Interestingly, none displayed significant variability within the diverse timespans and cadences of the available datasets. 

SN~2008bk is the only low-luminosity type IIP SN with multi-epoch progenitor observations, but these data have not featured in previous studies. The field of NGC~7793 containing SN~2008bk was repeatedly imaged as a part of Araucaria project \citep{Gieren05} with the 1.3m Warsaw telescope \citep{OGLE}. There are two sets of observations starting in mid-2004 that span 15 months in total, separated by $\sim$200\,d. 

Figure \ref{OGLElc} shows the $I$-band light curve of the progenitor. The average magnitude over the total time span of the observations is $m_{I}$=20.88$\pm$0.11, which is $\sim$0.38 mag brighter than that reported by \cite{2008bkMaund}. In the first group of observations from MJD=53250$-$53260, the source displays no significant variability ($\Delta m_I \leq0.1$ mag), barring a single epoch at MJD=53314.7 ($m_I$=21.13$\pm$0.19) that is marginally consistent with the above average value. In the second set of observations (MJD= 53615.7--53712.5) however, there appears to be a drop in brightness between MJD=53684--53685 of $\sim$0.6 mag from $m_I\sim20.8\pm0.13$ to $\sim21.4\pm0.24$\,mag. 

We find that there are 2 points that deviate from the mean value at the 3\,$\sigma$ level.
For a sample of 39 points, we would expect to observe none. Given the associated measurement uncertainties, and the apparent rapidity of the fluctuations, intrinsic source variability is at best tentative. Although our dataset does not formally permit us to invoke variability, we note that periodic and stochastic variability is a hallmark of RSGs, that is in some cases also accompanied by changes in temperature and local extinction \citep[e.g.,][]{Massey07,MontargesESO}. 

We remarked earlier that the low extinctions reported for the progenitors of other IIP SNe
were at odds with most RSG observations.
We speculate that if a large fraction of the IIP SNe with known progenitors arise from mergers \citep{Zapartas17}, then it is likely that any CSM dust will be destroyed or dissipated in this process.
Whether the merger product subsequently produces large quantities of dust inspite of the dramatic
alteration of its interior chemical profile, is a matter that remains to be investigated.

\begin{acknowledgements}
D. O'Neill acknowledges a DEL studentship award and FINCA for supporting a research visit. MF is supported by a Royal Society - Science Foundation Ireland University Research Fellowship. Support for JLP is provided in part by FONDECYT through the grant 1191038 and by ANID's Millennium Science Initiative through grant IC12\_009, awarded to The Millennium Institute of Astrophysics, MAS. The research leading to these results has received funding from the European Research Council (ERC) under the European Union’s Horizon 2020 research and innovation programme under grant agreement No 695099 (project CepBin).
\end{acknowledgements}

\bibliographystyle{aa}
\bibliography{2008bk}

\begin{thebibliography}{55}
\expandafter\ifx\csname natexlab\endcsname\relax\def\natexlab#1{#1}\fi

\bibitem[{{Alard}(2000)}]{Alard00}
{Alard}, C. 2000, \aaps, 144, 363

\bibitem[{{Alard} \& {Lupton}(1998)}]{Alard98}
{Alard}, C. \& {Lupton}, R.~H. 1998, \apj, 503, 325

\bibitem[{{Chiavassa} {et~al.}(2009){Chiavassa}, {Plez}, {Josselin}, \&
  {Freytag}}]{Chiavassa09}
{Chiavassa}, A., {Plez}, B., {Josselin}, E., \& {Freytag}, B. 2009, \aap, 506,
  1351

\bibitem[{{Chugai} \& {Utrobin}(2000)}]{97D}
{Chugai}, N.~N. \& {Utrobin}, V.~P. 2000, \aap, 354, 557

\bibitem[{{Davies} \& {Beasor}(2018)}]{IIpProDavies}
{Davies}, B. \& {Beasor}, E.~R. 2018, \mnras, 474, 2116

\bibitem[{{Davies} \& {Beasor}(2020)}]{DaviesRSGP}
{Davies}, B. \& {Beasor}, E.~R. 2020, \mnras, 493, 468

\bibitem[{{Davies} {et~al.}(2013){Davies}, {Kudritzki}, {Plez}, {Trager},
  {Lan{\c{c}}on}, {Gazak}, {Bergemann}, {Evans}, \& {Chiavassa}}]{Davies13}
{Davies}, B., {Kudritzki}, R.-P., {Plez}, B., {et~al.} 2013, \apj, 767, 3

\bibitem[{{Eggleton} {et~al.}(2011){Eggleton}, {Tout}, {Pols}, {Izzard},
  {Eldridge}, {Lesaffre}, {Stancliffe}, {Church}, \& {Lau}}]{STARS2}
{Eggleton}, P.~P., {Tout}, C., {Pols}, O., {et~al.} 2011, {STARS: A Stellar
  Evolution Code}, Astrophysics Source Code Library

\bibitem[{{Eldridge} {et~al.}(2017){Eldridge}, {Stanway}, {Xiao}, {McClelland},
  {Taylor}, {Ng}, {Greis}, \& {Bray}}]{BPASS2-1}
{Eldridge}, J.~J., {Stanway}, E.~R., {Xiao}, L., {et~al.} 2017, \pasa, 34, e058

\bibitem[{{Foreman-Mackey} {et~al.}(2013){Foreman-Mackey}, {Hogg}, {Lang}, \&
  {Goodman}}]{emcee}
{Foreman-Mackey}, D., {Hogg}, D.~W., {Lang}, D., \& {Goodman}, J. 2013, \pasp,
  125, 306

\bibitem[{{Fraser} {et~al.}(2014){Fraser}, {Maund}, {Smartt}, {Kotak},
  {Lawrence}, {Bruce}, {Valenti}, {Yuan}, {Benetti}, {Chen}, {Gal-Yam},
  {Inserra}, \& {Young}}]{2013ejPro}
{Fraser}, M., {Maund}, J.~R., {Smartt}, S.~J., {et~al.} 2014, \mnras, 439, L56

\bibitem[{{Gieren} {et~al.}(2005){Gieren}, {Pietrzynski}, {Bresolin},
  {Kudritzki}, {Minniti}, {Urbaneja}, {Soszynski}, {Storm}, {Fouque}, {Bono},
  {Walker}, \& {Garcia}}]{Gieren05}
{Gieren}, W., {Pietrzynski}, G., {Bresolin}, F., {et~al.} 2005, The Messenger,
  121, 23

\bibitem[{{Gustafsson} {et~al.}(2008){Gustafsson}, {Edvardsson}, {Eriksson},
  {J{\o}rgensen}, {Nordlund}, \& {Plez}}]{MARCS}
{Gustafsson}, B., {Edvardsson}, B., {Eriksson}, K., {et~al.} 2008, \aap, 486,
  951

\bibitem[{{Heger} {et~al.}(2003){Heger}, {Fryer}, {Woosley}, {Langer}, \&
  {Hartmann}}]{Heger}
{Heger}, A., {Fryer}, C.~L., {Woosley}, S.~E., {Langer}, N., \& {Hartmann},
  D.~H. 2003, \apj, 591, 288

\bibitem[{{Ivezic} {et~al.}(1999){Ivezic}, {Nenkova}, \&
  {Elitzur}}]{DustyManual}
{Ivezic}, Z., {Nenkova}, M., \& {Elitzur}, M. 1999, arXiv e-prints, astro

\bibitem[{{Jacobs} {et~al.}(2009){Jacobs}, {Rizzi}, {Tully}, {Shaya},
  {Makarov}, \& {Makarova}}]{2008bkDistTRGB}
{Jacobs}, B.~A., {Rizzi}, L., {Tully}, R.~B., {et~al.} 2009, \aj, 138, 332

\bibitem[{{Jerkstrand} {et~al.}(2018){Jerkstrand}, {Ertl}, {Janka},
  {M{\"u}ller}, {Sukhbold}, \& {Woosley}}]{AndersLowMass}
{Jerkstrand}, A., {Ertl}, T., {Janka}, H.~T., {et~al.} 2018, \mnras, 475, 277

\bibitem[{{Johnson} {et~al.}(2018){Johnson}, {Kochanek}, \&
  {Adams}}]{Johnson18}
{Johnson}, S.~A., {Kochanek}, C.~S., \& {Adams}, S.~M. 2018, \mnras, 480, 1696

\bibitem[{{Karachentsev} {et~al.}(2003){Karachentsev}, {Grebel}, {Sharina},
  {Dolphin}, {Geisler}, {Guhathakurta}, {Hodge}, {Karachentseva}, {Sarajedini},
  \& {Seitzer}}]{2008bkDistTRGBold}
{Karachentsev}, I.~D., {Grebel}, E.~K., {Sharina}, M.~E., {et~al.} 2003, \aap,
  404, 93

\bibitem[{{Kilpatrick} \& {Foley}(2018)}]{2017eawPro}
{Kilpatrick}, C.~D. \& {Foley}, R.~J. 2018, ArXiv e-prints
  [\eprint[arXiv]{1806.00348}]

\bibitem[{{Kochanek}(2020)}]{KochanekRSGP}
{Kochanek}, C.~S. 2020, \mnras, 493, 4945

\bibitem[{{Kochanek} {et~al.}(2017){Kochanek}, {Fraser}, {Adams}, {Sukhbold},
  {Prieto}, {M{\"u}ller}, {Bock}, {Brown}, {Dong}, {Holoien}, {Khan},
  {Shappee}, \& {Stanek}}]{16fq}
{Kochanek}, C.~S., {Fraser}, M., {Adams}, S.~M., {et~al.} 2017, \mnras, 467,
  3347

\bibitem[{{Kochanek} {et~al.}(2012){Kochanek}, {Khan}, \& {Dai}}]{12awCSM}
{Kochanek}, C.~S., {Khan}, R., \& {Dai}, X. 2012, \apj, 759, 20

\bibitem[{{Lan{\c{c}}on} {et~al.}(2007){Lan{\c{c}}on}, {Hauschildt}, {Ladjal},
  \& {Mouhcine}}]{Phoenix07}
{Lan{\c{c}}on}, A., {Hauschildt}, P.~H., {Ladjal}, D., \& {Mouhcine}, M. 2007,
  \aap, 468, 205

\bibitem[{{Levesque} {et~al.}(2005){Levesque}, {Massey}, {Olsen}, {Plez},
  {Josselin}, {Maeder}, \& {Meynet}}]{Levesque05}
{Levesque}, E.~M., {Massey}, P., {Olsen}, K.~A.~G., {et~al.} 2005, \apj, 628,
  973

\bibitem[{{Li} {et~al.}(2011){Li}, {Leaman}, {Chornock}, {Filippenko},
  {Poznanski}, {Ganeshalingam}, {Wang}, {Modjaz}, {Jha}, {Foley}, \&
  {Smith}}]{LiRates}
{Li}, W., {Leaman}, J., {Chornock}, R., {et~al.} 2011, \mnras, 412, 1441

\bibitem[{{Li} {et~al.}(2006){Li}, {Van Dyk}, {Filippenko}, {Cuillandre},
  {Jha}, {Bloom}, {Riess}, \& {Livio}}]{Pro05csLi}
{Li}, W., {Van Dyk}, S.~D., {Filippenko}, A.~V., {et~al.} 2006, \apj, 641, 1060

\bibitem[{{Lisakov} {et~al.}(2017){Lisakov}, {Dessart}, {Hillier}, {Waldman},
  \& {Livne}}]{2008bkLisakov}
{Lisakov}, S.~M., {Dessart}, L., {Hillier}, D.~J., {Waldman}, R., \& {Livne},
  E. 2017, \mnras, 466, 34

\bibitem[{{Martinez} {et~al.}(2020){Martinez}, {Bersten}, {Anderson},
  {Gonz{\'a}lez-Gait{\'a}n}, {F{\"o}rster}, \& {Folatelli}}]{Martinez20}
{Martinez}, L., {Bersten}, M.~C., {Anderson}, J.~P., {et~al.} 2020, \aa,
  arXiv:2008.05572

\bibitem[{{Massey} {et~al.}(2007){Massey}, {Levesque}, {Olsen}, {Plez}, \&
  {Skiff}}]{Massey07}
{Massey}, P., {Levesque}, E.~M., {Olsen}, K.~A.~G., {Plez}, B., \& {Skiff},
  B.~A. 2007, \apj, 660, 301

\bibitem[{{Mattila} {et~al.}(2010){Mattila}, {Smartt}, {Maund}, {Benetti}, \&
  {Ergon}}]{2008bkDisap}
{Mattila}, S., {Smartt}, S., {Maund}, J., {Benetti}, S., \& {Ergon}, M. 2010,
  arXiv e-prints, arXiv:1011.5494

\bibitem[{{Mattila} {et~al.}(2008){Mattila}, {Smartt}, {Eldridge}, {Maund},
  {Crockett}, \& {Danziger}}]{2008bkVLT}
{Mattila}, S., {Smartt}, S.~J., {Eldridge}, J.~J., {et~al.} 2008, \apjl, 688,
  L91

\bibitem[{{Maund}(2017)}]{2008bkPop}
{Maund}, J.~R. 2017, \mnras, 469, 2202

\bibitem[{{Maund} {et~al.}(2014){Maund}, {Mattila}, {Ramirez-Ruiz}, \&
  {Eldridge}}]{2008bkMaund}
{Maund}, J.~R., {Mattila}, S., {Ramirez-Ruiz}, E., \& {Eldridge}, J.~J. 2014,
  \mnras, 438, 1577

\bibitem[{{Maund} {et~al.}(2005){Maund}, {Smartt}, \& {Danziger}}]{05csPro}
{Maund}, J.~R., {Smartt}, S.~J., \& {Danziger}, I.~J. 2005, \mnras, 364, L33

\bibitem[{{Montargès}(2020)}]{MontargesESO}
{Montargès}, M. 2020, https://www.eso.org/public/news/eso2003/

\bibitem[{{Nomoto} {et~al.}(2013){Nomoto}, {Kobayashi}, \&
  {Tominaga}}]{Nomoto13}
{Nomoto}, K., {Kobayashi}, C., \& {Tominaga}, N. 2013, \araa, 51, 457

\bibitem[{{O'Neill} {et~al.}(2019){O'Neill}, {Kotak}, {Fraser}, {Sim},
  {Benetti}, {Smartt}, {Mattila}, {Ashall}, {Callis}, {Elias-Rosa},
  {Gromadzki}, \& {Prentice}}]{18aoqPro}
{O'Neill}, D., {Kotak}, R., {Fraser}, M., {et~al.} 2019, \aap, 622, L1

\bibitem[{{Pastorello} {et~al.}(2009){Pastorello}, {Valenti}, {Zampieri},
  {Navasardyan}, {Taubenberger}, {Smartt}, {Arkharov}, {B{\"a}rnbantner},
  {Barwig}, {Benetti}, {Birtwhistle}, {Botticella}, {Cappellaro}, {Del
  Principe}, {di Mille}, {di Rico}, {Dolci}, {Elias-Rosa}, {Efimova},
  {Fiedler}, {Harutyunyan}, {H{\"o}flich}, {Kloehr}, {Larionov}, {Lorenzi},
  {Maund}, {Napoleone}, {Ragni}, {Richmond}, {Ries}, {Spiro}, {Temporin},
  {Turatto}, \& {Wheeler}}]{2005csPast}
{Pastorello}, A., {Valenti}, S., {Zampieri}, L., {et~al.} 2009, \mnras, 394,
  2266

\bibitem[{{Pietrzy{\'n}ski} {et~al.}(2010){Pietrzy{\'n}ski}, {Gieren}, {Hamuy},
  {Pignata}, {Soszy{\'n}ski}, {Udalski}, {Walker}, {Fouqu{\'e}}, {Bresolin},
  {Kudritzki}, {Garcia-Varela}, {Szewczyk}, {Szyma{\'n}ski}, {Kubiak}, \&
  {Wyrzykowski}}]{2008bkDist}
{Pietrzy{\'n}ski}, G., {Gieren}, W., {Hamuy}, M., {et~al.} 2010, \aj, 140, 1475

\bibitem[{{Plez}(2011)}]{Plez}
{Plez}, B. 2011, in Journal of Physics Conference Series, Vol. 328, Journal of
  Physics Conference Series, 012005

\bibitem[{{Poelarends} {et~al.}(2008){Poelarends}, {Herwig}, {Langer}, \&
  {Heger}}]{ECSNPoel}
{Poelarends}, A.~J.~T., {Herwig}, F., {Langer}, N., \& {Heger}, A. 2008, \apj,
  675, 614

\bibitem[{{Pumo} {et~al.}(2009){Pumo}, {Turatto}, {Botticella}, {Pastorello},
  {Valenti}, {Zampieri}, {Benetti}, {Cappellaro}, \& {Patat}}]{Pumo09}
{Pumo}, M.~L., {Turatto}, M., {Botticella}, M.~T., {et~al.} 2009, \apjl, 705,
  L138

\bibitem[{{Quataert} \& {Shiode}(2012)}]{Quataert12}
{Quataert}, E. \& {Shiode}, J. 2012, \mnras, 423, L92

\bibitem[{{Schlafly} \& {Finkbeiner}(2011)}]{SF2011}
{Schlafly}, E.~F. \& {Finkbeiner}, D.~P. 2011, \apj, 737, 103

\bibitem[{{Smartt}(2015)}]{Smartt15}
{Smartt}, S.~J. 2015, \pasa, 32, e016

\bibitem[{{Smartt} {et~al.}(2009){Smartt}, {Eldridge}, {Crockett}, \&
  {Maund}}]{RSGP}
{Smartt}, S.~J., {Eldridge}, J.~J., {Crockett}, R.~M., \& {Maund}, J.~R. 2009,
  \mnras, 395, 1409

\bibitem[{{Smith}(2014)}]{Smith14}
{Smith}, N. 2014, \araa, 52, 487

\bibitem[{{Smith} \& {Arnett}(2014)}]{SmithArnett14}
{Smith}, N. \& {Arnett}, W.~D. 2014, \apj, 785, 82

\bibitem[{{Udalski} {et~al.}(1992){Udalski}, {Szymanski}, {Kaluzny}, {Kubiak},
  \& {Mateo}}]{OGLE}
{Udalski}, A., {Szymanski}, M., {Kaluzny}, J., {Kubiak}, M., \& {Mateo}, M.
  1992, \actaa, 42, 253

\bibitem[{{Van Dyk} {et~al.}(2012){Van Dyk}, {Davidge}, {Elias-Rosa},
  {Taubenberger}, {Li}, {Levesque}, {Howerton}, {Pignata}, {Morrell}, {Hamuy},
  \& {Filippenko}}]{2008bkSVD}
{Van Dyk}, S.~D., {Davidge}, T.~J., {Elias-Rosa}, N., {et~al.} 2012, \aj, 143,
  19

\bibitem[{{Walmswell} \& {Eldridge}(2012)}]{RSGPdust}
{Walmswell}, J.~J. \& {Eldridge}, J.~J. 2012, \mnras, 419, 2054

\bibitem[{{Zampieri} {et~al.}(2003){Zampieri}, {Pastorello}, {Turatto},
  {Cappellaro}, {Benetti}, {Altavilla}, {Mazzali}, \& {Hamuy}}]{Zamp03}
{Zampieri}, L., {Pastorello}, A., {Turatto}, M., {et~al.} 2003, \mnras, 338,
  711

\bibitem[{{Zapartas} {et~al.}(2017){Zapartas}, {de Mink}, {Izzard}, {Yoon},
  {Badenes}, {G{\"o}tberg}, {de Koter}, {Neijssel}, {Renzo}, {Schootemeijer},
  \& {Shrotriya}}]{Zapartas17}
{Zapartas}, E., {de Mink}, S.~E., {Izzard}, R.~G., {et~al.} 2017, \aap, 601,
  A29

\bibitem[{{Zgirski} {et~al.}(2017){Zgirski}, {Gieren}, {Pietrzy{\'n}ski},
  {Karczmarek}, {Gorski}, {Wielgorski}, {Narloch}, {Graczyk}, {Kudritzki}, \&
  {Bresolin}}]{2008bkdist2}
{Zgirski}, B., {Gieren}, W., {Pietrzy{\'n}ski}, G., {et~al.} 2017, \apj, 847,
  88

\end{thebibliography}

\begin{appendix}
\section{additional figures and tables}
\begin{figure*}
   \centering
   \includegraphics[width=0.9\hsize]{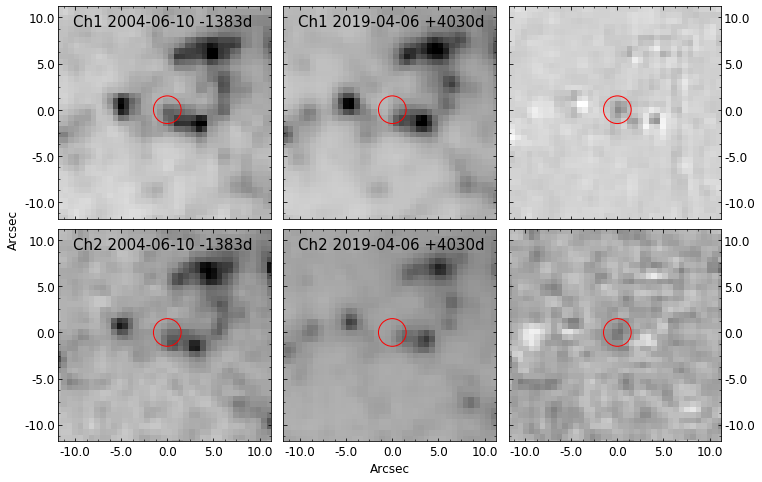}
      \caption{\textbf{Left:} The pre-SN images showing the progenitor (red) circle in channels 1 (top) and 2 (bottom). \textbf{Middle:} The post-explosion images in channels 1 and 2. \textbf{Right:} Difference images (post-explosion $-$ pre-explosion image). There is a source visible at the progenitor position in both channels. We detect it with a SNR of $\sim$10 ($\gg3\sigma$) in channel 1, and a SNR of $\sim$4 ($\sim 3\sigma$) in channel 2.
      }
         \label{ch1}
   \end{figure*}
   
\begin{figure*}
   \centering
   \includegraphics[width=0.7\hsize]{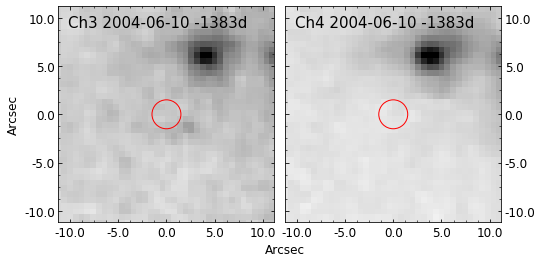}
      \caption{Pre-explosion images showing the progenitor position (red circle) in IRAC channels 3 (left) and 4 (right). Upper limits on the flux at this location are given in Table \ref{IRACtable}.}
         \label{ch34}
   \end{figure*}

\begin{figure}
   \centering
   \includegraphics[width=1\hsize]{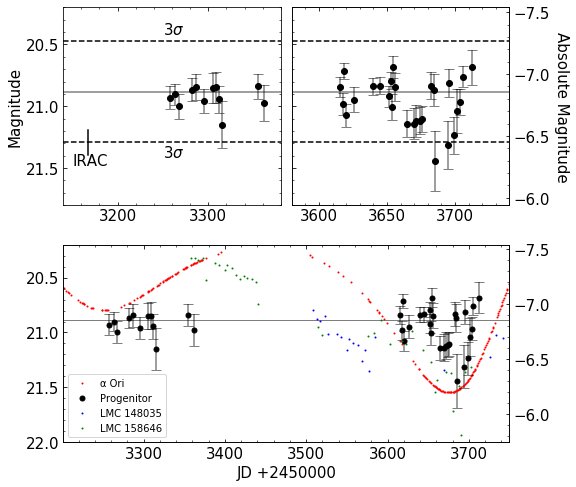}
      \caption{\textbf{Top:} The $I$ band light curve of the progenitor of SN~2008bk. The solid line denotes the weighted mean average magnitude for the dataset. The dashed lines represent the 3$\sigma$ variation for the dataset. The values have only been corrected for Milky Way reddening. The vertical line shows the epochs at which the IRAC data were observed. \textbf{Bottom:} The lightcurve of the progenitor alongside three other RSG stars. The $\alpha$ Ori V-band light curve data was obtained from AAVSO\protect\footnote{https://www.aavso.org/} while the $V$-band data for the LMC RSGs are from ASAS\protect\footnote{http://www.astrouw.edu.pl/asas/?page=main}. There is limited $I$-band data for the objects. The brightness of the sample RSGs are scaled such that the mean magnitude over the dataset is the same as that of the progenitor, and shifted in time such that the minimum in the light curve matches that of the progenitor. The progenitor of SN~2008bk does not show the slow variations displayed by well-known Galactic and other LMC RSGs.}
         \label{OGLElc}
\end{figure}

%%%%%%%%%%%%%%%%%%%%%%

\begin{table}[!h]
\begin{tabular}{c c}
\hline\hline                 % inserts double horizontal lines
\tabularnewline[-0.25cm]
Channel & Magnitude \\
 & (mag) 
\tabularnewline% table heading 
\hline                        % inserts single horizontal line
\tabularnewline[-0.25cm]
1 & 17.85 (0.10)\\
2 & 18.28 (0.25)\\
3 & >15.87 \\
4 & >15.12 \\

\hline\hline %inserts single line
\tabularnewline[-0.25cm]

\end{tabular}
\caption{Summary of the pre-explosion IRAC data used in this paper. All images were taken at 2004-06-10 (JD=2453166.5) (Kennicut et al. Proposal ID: 154). 2018-04-06 template image: Kasliwal et al. Proposal ID: 14089.}
\label{IRACtable}  
\end{table}

\begin{table}
\begin{tabular}{c c c c }
\hline\hline                 % inserts double horizontal lines
\tabularnewline[-0.25cm]
Date & JD & Phase & Magnitude \\
 & (+2450000) & (d) & (mag) \tabularnewline% table heading 
\hline                        % inserts single horizontal line
\tabularnewline[-0.25cm]
2004-09-09 &	3257.8	&	-1292.2	&	20.910	(0.094	)\\
2004-09-15 &	3263.8	&	-1286.2	&	20.882	(0.084	)\\
2004-09-19 &	3267.8	&	-1282.2	&	20.979	(0.101	)\\
2004-10-03 &	3281.7	&	-1268.3	&	20.845	(0.092	)\\
2004-10-08 &	3286.7	&	-1263.3	&	20.824	(0.103	)\\
2004-10-17 &	3295.7	&	-1254.3	&	20.937	(0.100	)\\
2004-10-27 &	3305.6	&	-1244.4	&	20.832	(0.119	)\\
2004-10-30 &	3308.7	&	-1241.3	&	20.827	(0.123	)\\
2004-11-02 &	3311.6	&	-1238.4	&	20.924	(0.112	)\\
2004-11-05 &	3314.7	&	-1235.3	&	21.129	(0.191	)\\
2004-12-15 &	3354.6	&	-1195.4	&	20.820	(0.102	)\\
2004-12-22 &	3361.5	&	-1188.5	&	20.954	(0.146	)\\
2005-09-02 &	3615.7	&	-934.3	&	20.825	(0.082	)\\
2005-09-04 &	3617.7	&	-932.3	&	20.959	(0.088	)\\
2005-09-05 &	3618.7	&	-931.3	&	20.693	(0.067	)\\
2005-09-06 &	3619.8	&	-930.2	&	21.054	(0.091	)\\
2005-09-13 &	3625.9	&	-924.1	&	20.927	(0.102	)\\
2005-09-26 &	3639.7	&	-910.3	&	20.819	(0.069	)\\
2005-10-01 &	3644.7	&	-905.3	&	20.814	(0.068	)\\
2005-10-08 &	3651.6	&	-898.4	&	20.899	(0.084	)\\
2005-10-09 &	3652.6	&	-897.4	&	20.774	(0.069	)\\
2005-10-10 &	3653.6	&	-896.4	&	20.986	(0.109	)\\
2005-10-11 &	3654.6	&	-895.4	&	20.663	(0.089	)\\
2005-10-12 &	3655.6	&	-894.4	&	20.827	(0.113	)\\
2005-10-21 &	3664.6	&	-885.4	&	21.121	(0.107	)\\
2005-10-26 &	3669.6	&	-880.4	&	21.120	(0.123	)\\
2005-10-28 &	3671.7	&	-878.3	&	21.099	(0.127	)\\
2005-10-31 &	3674.6	&	-875.4	&	21.097	(0.108	)\\
2005-11-01 &	3675.6	&	-874.4	&	21.084	(0.108	)\\
2005-11-08 &	3682.5	&	-867.5	&	20.814	(0.109	)\\
2005-11-10 &	3684.6	&	-865.4	&	20.849	(0.128	)\\
2005-11-11 &	3685.6	&	-864.4	&	21.422	(0.242	)\\
2005-11-20 &	3694.7	&	-855.3	&	21.292	(0.194	)\\
2005-11-21 &	3695.5	&	-854.5	&	20.792	(0.111	)\\
2005-11-25 &	3699.5	&	-850.5	&	21.215	(0.150	)\\
2005-11-27 &	3701.6	&	-848.4	&	21.019	(0.175	)\\
2005-11-29 &	3703.6	&	-846.4	&	20.947	(0.105	)\\
2005-12-01 &	3705.6	&	-844.4	&	20.741	(0.084	)\\
2005-12-08 &	3712.5	&	-837.5	&	20.663	(0.142	)\\
\hline\hline %inserts single line
\tabularnewline[-0.25cm]

\end{tabular}
\caption{Table of the OGLE $I$-band data. These values have not been corrected for reddening. The phase shown is with respect to the estimated explosion epoch of 2008-03-24 \citep[JD=2454550,][]{2008bkSVD}.
}
\label{OGLEdata}  
\end{table}

\end{appendix}

\end{document}